\documentclass[preprint, 12p]{elsarticle}
\usepackage{amssymb}
\usepackage{amsmath}
\usepackage{graphicx}
\usepackage{dcolumn}
\usepackage{bm}
\usepackage{hyperref}
\usepackage[nodots,nocompress]{numcompress}
\usepackage{multirow}
\usepackage{array}
\usepackage[usenames,dvipsnames]{color}
\usepackage{listings}
\journal{Computers \& Fluids}
\begin{document}

\begin{frontmatter}


\title{Benchmarking and scaling studies of a pseudospectral code Tarang for turbulence simulations}

\author[physiitk]{Mahendra K. Verma\corref{cor1}}
\ead{mkv@iitk.ac.in}

\author[physiitk]{Anando Chaterjee}

\author[cdac]{Supriyo Paul}

\author[mechiitk]{Sandeep Reddy}

\author[physiitk]{Rakesh K. Yadav}

\author[physiitk]{Mani Chandra} 

\author[mechkaust]{Ravi Samtaney} 
\cortext[cor1]{Corresponding author}

\address[physiitk]{Department of Physics, Indian Institute of Technology Kanpur, India 208016}
\address[cdac]{Computational Fluid Dynamics Team, Centre for Development of Advanced Computing, Pune, India 411007}
\address[mechiitk]{Department of Mechanical Engineering, Indian Institute of Technology Kanpur, India 208016}
\address[mechkaust]{Department of Mechanical Engineering, King Abdullah University of Science and Technology Thuwal, Kingdom of Saudi Arabia 23955-6900}


\begin{abstract}
Tarang is a general-purpose pseudospectral parallel code for simulating flows involving fluids, magnetohydrodynamics, and Rayleigh-B\'{e}nard convection in turbulence and instability regimes.  In this paper we present code validation and benchmarking results of Tarang.  We performed our simulations on $1024^3$, $2048^3$, and $4096^3$ grids using the {\em HPC system} of IIT Kanpur and {\em Shaheen} of KAUST.  We observe good ``weak" and ``strong" scaling for Tarang on these systems.
\end{abstract}


\end{frontmatter}

\section{\label{sec:Introduction}  Introduction}
A typical fluid flow is random or chaotic in the turbulent and instability regimes.  Therefore we need to employ accurate numerical schemes for simulating such flows.  A pseudospectral algorithm~\cite{Boyd:book,Canuto:book} is one of the  most accurate methods for solving fluid flows, and it is employed for performing direct numerical simulations of turbulent flows, as well as for critical applications like weather predictions and climate modelling.   Yokokawa et al.~\cite{Kaneda:techrep, Kaneda:Pof2003}, Donzis et al.~\cite{Donzis}, and Pouquet et al.~\cite{Pouquet} have performed spectral simulations on some of the largest grids (e.g., $4096^3$).

We have developed a general-purpose flow solver named Tarang (synonym for waves in Sanskrit) for turbulence and instability studies.  Tarang is a parallel and modular code written in object-oriented language C++.  Using Tarang, we can solve incompressible flows involving pure fluid, Rayleigh-B\'{e}nard convection, passive and active scalars, magnetohydrodynamics, liquid-metals, etc.  Tarang is an open-source code and it can be downloaded from \url{http://turbulence.phy.iitk.ac.in}.  In this paper we will describe some details of the code, scaling results, and code validation performed on Tarang.

\section{Salient features of Tarang}
The basic steps of Tarang follow the standard procedure of pseudospectral method~\cite{Boyd:book,Canuto:book}.  The Navier-Stokes and related equations are numerically solved given an initial condition of the fields.  The fields are time-stepped using one of the time integrators.   The nonlinear terms, e.g. ${\bf u\cdot \nabla u}$,  transform to convolutions in the spectral space, which are very expensive to compute.  Orszag devised a clever scheme to compute the convolution in an efficient manner using Fast Fourier Transforms (FFT)~\cite{Boyd:book,Canuto:book}.   In this scheme, the fields are transformed from the Fourier space to the real space, multiplied with each other, and then transformed back to the Fourier space.  Note that the spectral transforms could involve Fourier functions, sines and cosines, Chebyshev polynomials, spherical harmonics, or a combination of these functions depending on the boundary conditions.  For details the reader is referred to standard references, e.g., the books by Boyd~\cite{Boyd:book} and Canuto et al.~\cite{Canuto:book}.  Some of the specific choices made in Tarang are as follow:

\begin{enumerate}[(a)]
\item In the turbulent regime, the two relevant time scales, the large-eddy turnover time and the small-scale viscous time, are very different (order of magnetic apart).  To handle this feature, we use the ``exponential trick" that absorbs the viscous term using a change of variable~\cite{Canuto:book}.
\item We use the fourth-order Runge-Kutta scheme for time stepping.  The code however has an option to use the Euler and the second-order Runge-Kutta schemes as well.
\item The code provides an option for dealiasing the fields.  The 3/2 rule is used for dealiasing~\cite{Canuto:book}.
\item The wavenumber components $k_i$ are
\begin{equation}
k_i = \frac{2\pi}{L_i} n_i
\end{equation}
where $L_i$ is the box dimension in the $i$-th direction, and $n_i$ is an integer.  We use parameters
\begin{equation}
\mathrm{kfactor}_i = \frac{2\pi}{L_i}
\end{equation}
to control the box size, especially for Rayleigh-B\'{e}nard convection.  Note that typical spectral codes take $ \mathrm{kfactor}_i = 1$, or $k_i = n_i$.
\end{enumerate}


The parallel implementation of Tarang involved parallelization of the spectral transforms and the input-output operations, as described below.

\section{\label{sec:Pstrategy} Parallelization Strategy}
A pseudospectral code involves forward and inverse transforms between the spectral and real space.  In a typical pseudospectral code, these operations take approximately 80\% of the total time.  Therefore, we use one of the most efficient parallel FFT routines, FFTW (Fastest Fourier Transform in the West)~\cite{Frigo:IEEE2005}, in Tarang. We adopt FFTW's strategy for dividing the arrays.   If $p$ is the number of available processors, we divide each of the arrays into $p$ ``slabs".  For example, a complex array $A(N_1, N_2, N_3/2+1)$ is split into $A(N_1/p, N_2, N_3/2+1)$ segments, each of which is handled by a single processor.  This division  is called ``slab decomposition".  The other time-consuming tasks in Tarang are the input and output (I/O) operations of large data sets, and the element-by-element multiplication of arrays.  The data sets in Tarang are massive, for example, the data size of a $4096^3$ fluid  simulation  is of the order of 1.5 terabytes.   For I/O operations, we use an efficient and parallel library named HDF5 (Hierarchical Data Format-5).  The third operation, element-by-element multiplication of arrays, is handled by individual processors in a straightforward manner. 

Tarang has been organized in a modular fashion, so the spectral transforms and I/O operations were easily parallelized.   For a periodic-box, we use the parallel FFTW library itself.  However, for the mixed transforms (e.g., sine transform along $x$, and Fourier transform along $yz$ plane), we parallelize the transforms ourselves using one- and two-dimensional FFTW transforms. 

An important aspect of any parallel simulation code is its scalability. We tested the scaling of FFTW and Tarang by performing simulations on $1024^3$, $2048^3$, and $4096^3$ grids with variable number of processors.  The simulations were performed on the {\em HPC system} of IIT Kanpur and  {\em Shaheen} supercomputer of King Abdullah University of Science and Technology (KAUST).   The {\em HPC system}  has 368 compute nodes connected via a 40 Gbps Qlogic Infiniband switch with each node containing dual Intel Xeon Quadcore C5570 processor and 48GB of RAM.   Its peak performance (Rpeak) is approximately 34 teraflops (tera floating point operations per second).  {\em Shaheen} on the other hand is a 16-rack IBM BlueGene/P system with  65536 cores and 65536 GB of RAM.  {\em Shaheen}'s peak performance is approximately 222 teraflops.

\begin{figure}[htbp]
\begin{center}
\includegraphics[scale=0.5]{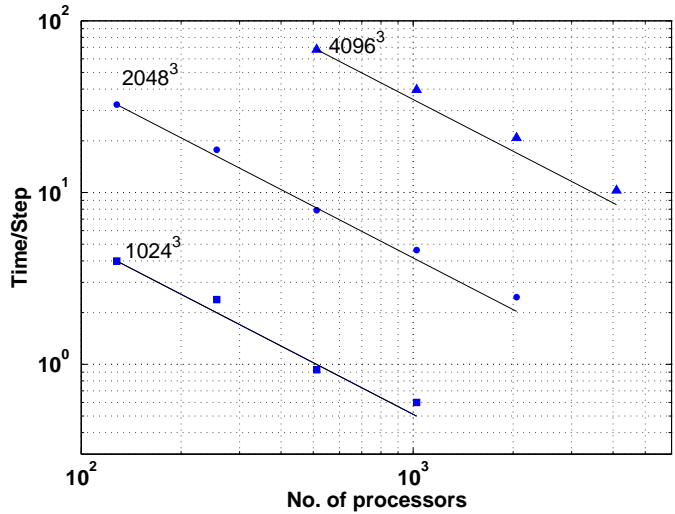}
\end{center}
\caption{Scaling of parallel FFT on {\em Shaheen} for $1024^3, 2048^3$ and $4096^3$ grids with single precision computation. The straight lines represent the ideal linear scaling.}  
\label{fig:bg_slab_fft}
\end{figure}
 
For parallel FFT with slab decomposition, we compute the time taken per step (forward+backward transform) on {\em Shaheen} for several large $N^3$ grids.  The results displayed in Fig.~\ref{fig:bg_slab_fft} demonstrate an approximate linear scaling (called ``strong scaling").    Using the fact that each forward plus inverse FFT involves $5 N^3 \log N^3$ operations for single precision computations~\cite{Frigo:IEEE2005}, the average FFT performance per core on {\em Shaheen} is approximately 0.3 gigaflops, which is only 8\% of its peak performance.  Similar efficiency is observed for the {\em HPC system} as well, whose cores have rating of approximately 12 gigaflops.  The aforementioned loss of efficiency is consistent with the other FFT libraries, e.g, p3dfft~\cite{P3dfft:online}.  Also note that an increase in the data size and number of processors (resources) by a same amount takes approximately the same time (see Fig.~\ref{fig:bg_slab_fft}).  For example, FFT of a $1024^3$ array using 128 processors, as well as that of a $2048^3$ array on 1024 processors, takes approximately 4 seconds.  Thus our implementation of FFT shows good ``weak scaling" as well.

\begin{figure}[htbp]
\begin{center}
\includegraphics[scale=0.6]{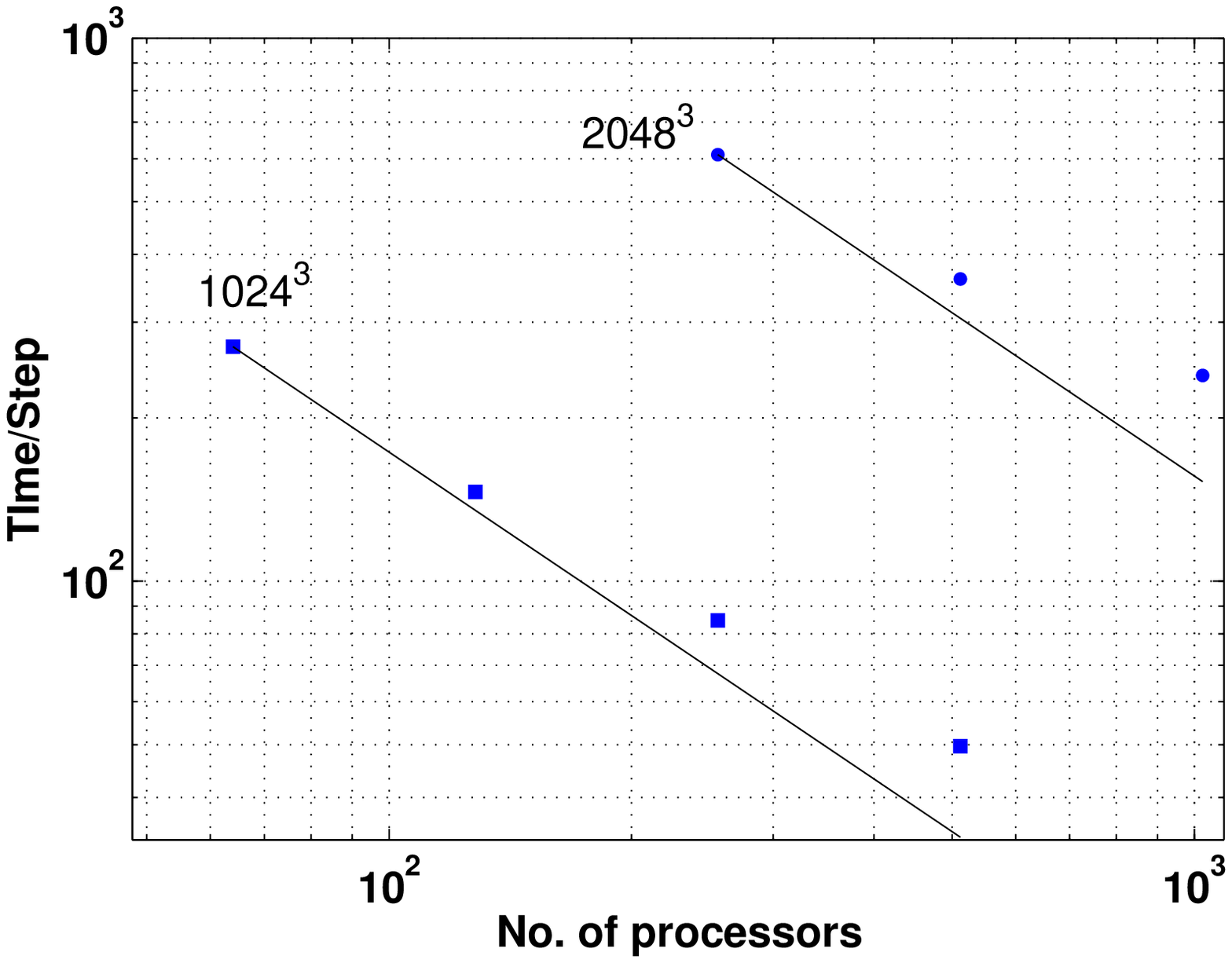}
\end{center}
\caption{Scaling of Tarang's fluid solver on {\em Shaheen} for $1024^3$ and $2048^3$ grids with single precision computation. The straight lines represent the ideal linear scaling.}  
 \label{fig:BG_Tarang_Fluid}
 \end{figure}
 
\begin{figure}[htbp]
\begin{center}
\includegraphics[scale=0.6]{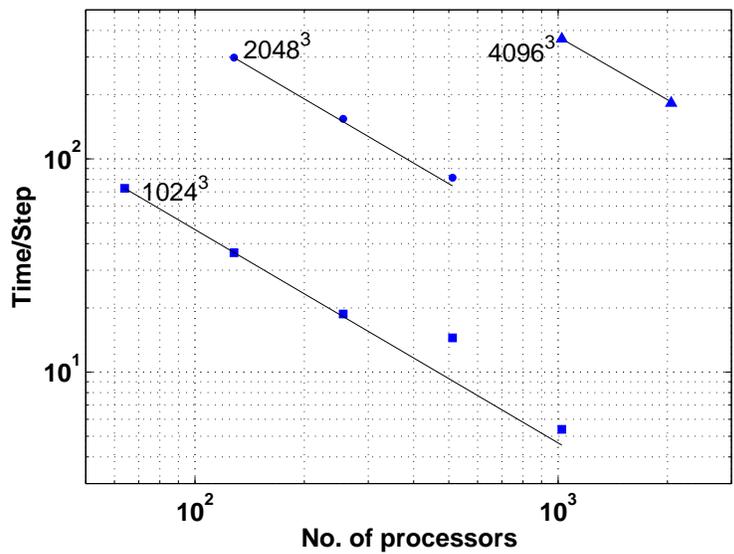}
\end{center}
\caption{Scaling of Tarang's fluid solver on the {\em HPC system} of IIT Kanpur for $1024^3$, $2048^3$, and $4096^3$ grids with single precision computation.   The straight lines represent the ideal linear scaling.}  
\label{fig:HPC_Tarang_Fluid}
\end{figure}
 
\begin{figure}[htbp]
\begin{center}
\includegraphics[scale=0.6]{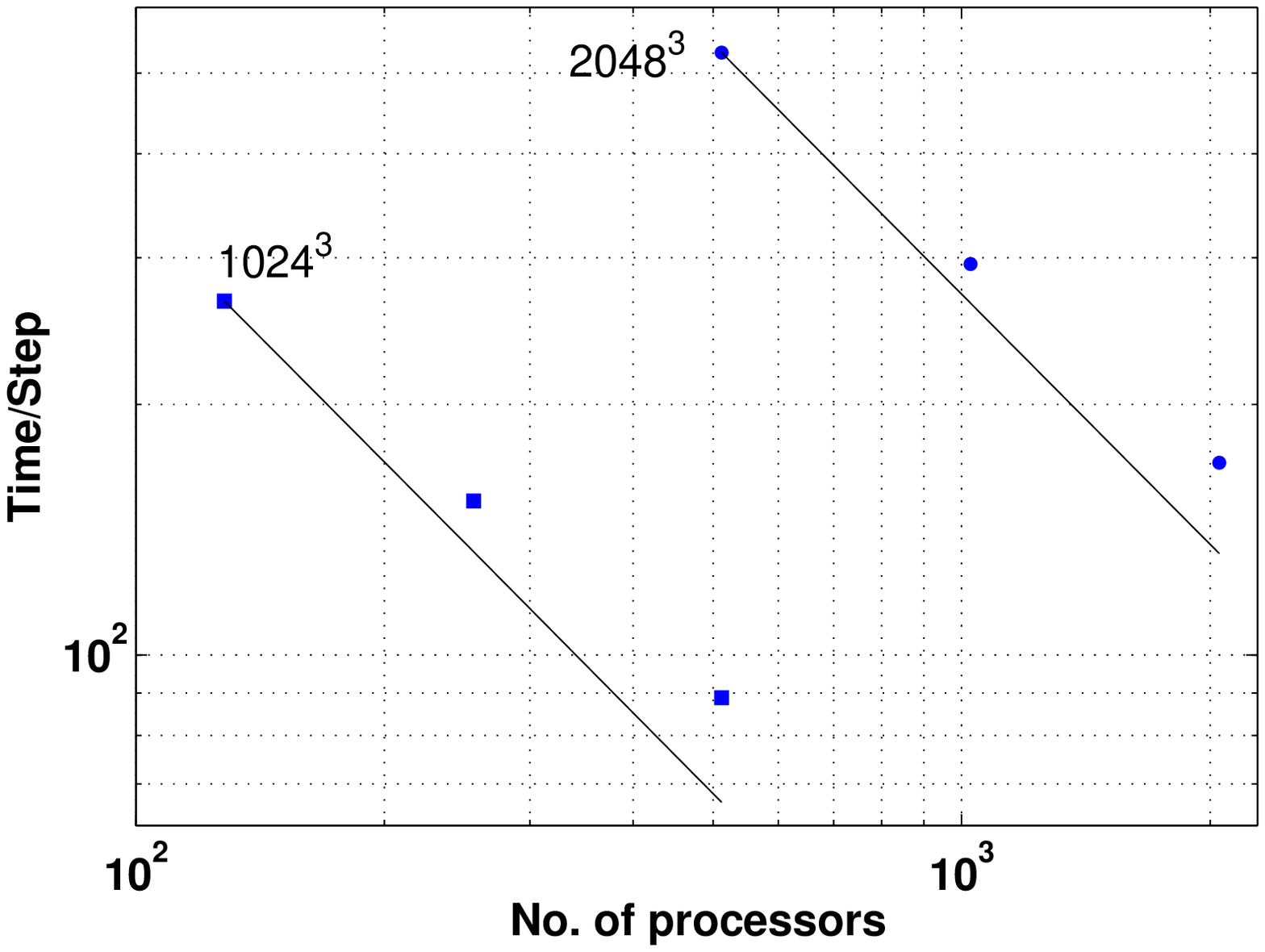}
\end{center}
\caption{Scaling of Tarang's magnetohydrodynamic (MHD) solver  on {\em Shaheen} for $1024^3$ and $2048^3$ grids  with single precision computation. The straight lines represent the ideal linear scaling.  }  
\label{fig:BG_Tarang_MHD_GNU}
\end{figure}
 
We also test the scaling of Tarang on  {\em Shaheen} and the {\em HPC system}.
Figs.~\ref{fig:BG_Tarang_Fluid} and \ref{fig:HPC_Tarang_Fluid} exhibit the scaling results of fluid simulations performed on these systems. Fig.~\ref{fig:BG_Tarang_MHD_GNU} shows the scaling results for magnetohydrodynamics (MHD) simulation on Shaheen.  These plots demonstrate {\em strong scaling} of Tarang, consistent with the aforementioned FFT scaling.  Sometimes we observe a small loss of efficiency when  $N = p$. We also observe approximate {\em weak scaling} for Tarang on both {\em Shaheen} and the {\em HPC system}.

A critical limitation of the ``slab decomposition" is that the number of processor cannot be more than $N_1$. This limitation can be overcome in a new scheme called ``pencil decomposition" in which the array $A(N_1, N_2, N_3/2+1)$ is split into $A(N_1/p_1, N_2/p_2, N_3/2+1)$ pencils where the total number of processors $p=p_1 \times p_2$~\cite{P3dfft:online}.  We are in the process of implementing ``pencil decomposition" on Tarang. In this paper we will focus only on the ``slab decomposition".

After the above discussion on parallelization of the code, we will discuss code validation, and time and space complexities for simulations of fluid turbulence, Rayleigh-B\'{e}nard convection, and magnetohydrodynamic turbulence.

\section{Fluid turbulence}
The governing equations for incompressible fluid turbulence are
\begin{eqnarray}
\partial_{t}{\bf u} + ({\bf u}\cdot\nabla){\bf u} &=& -\nabla{p}+ \nu\nabla^2 {\bf u} + {\bf F}^u, \label{eq:NS}\\
\nabla \cdot {\bf u} & = & 0, \label{eq:continuity} 
\end{eqnarray}
where ${\bf u}$ is the velocity field, $p$ is the pressure field, $\nu$ is the kinematic viscosity, and ${\bf F}^u$ is the external forcing.    For studies on homogeneous and isotropic turbulence, simulations are performed on high-resolution grids (e.g, $2048^3, 4096^3$) with a periodic boundary condition.   The resolution requirement is stringent due to $N \sim Re^{3/4}$ relation; for $Re=10^5$,  the required grid resolution is approximately $5600^3$, which is quite challenging even for modern supercomputers.  
 
Regarding the space complexity of a forced fluid turbulence simulation, Tarang requires 15 arrays (for ${\bf u(k), u(r)}, {\bf F}^u {\bf (k), nlin(k)}$, and three temporary arrays), which translates to approximately 120 gigabytes (8 terabytes) of  memory for $1024^3$  ($4096^3$) double-precision computations.  Here {\bf k} and {\bf r} represent the wavenumbers and the real space coordinates respectively.  The requirement is halved for a simulation with single precision.     Regarding the time requirement, each numerical step of the fourth-order Runge-Kutta (RK4) scheme requires $9\times4$ FFT operations.  The factor 9 is due to the 3 inverse and 6 forward transforms performed for each of the four RK4 iterates.  Therefore, for every time step, all the FFT operations require $36 \times 2.5\times N^3 \log_2(N^3)$ multiplications for a single precision simulation~\cite{Frigo:IEEE2005}, which translates to approximately 2.9 (185) tera floating-point operations for $1024^3$ ($4096^3$) grids.   The number of operations for double-precision computation is twice of the above estimate.  On 128 processors on {\it HPC system}, a fluid simulation with single-precision takes approximately 36 seconds (see Fig.~\ref{fig:HPC_Tarang_Fluid}), which corresponds to per core performance of approximately 0.68 gigaflops.  This is only 6\% of the peak performance of the cores, which is consistent with the efficiency of FFT operations discussed in Section~\ref{sec:Pstrategy}.  Also note that the solver also involves other operations, e.g., element-by-element array multiplication, but these operations take only a small fraction of the total time. 

We can also estimate the total time required to perform a $4096^3$ fluid simulation. A typical fluid turbulence would require 5 eddy turnover time with $dt \approx 5\times 10^{-4}$, which corresponds to $10^4$ time steps for the simulations.  So the total floating point operations required for this single-precision simulation is $185\times 10^4$ tera floating-point operations for the FFT itself.   Assuming 5\% efficiency for FFT, and FFTs share being 80\% of the total time,  the aforementioned fluid simulation will take approximately 128 hours on a 100 teraflop cluster.  

\begin{figure}[htbp]
\begin{center}
\includegraphics[scale=0.35]{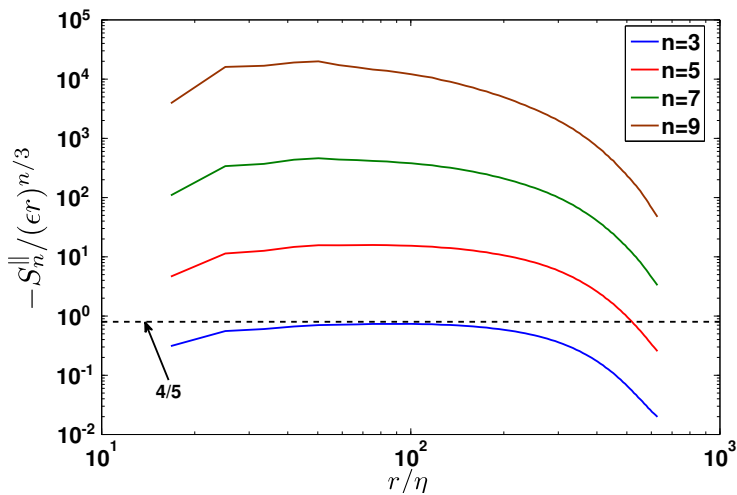}
\end{center}
\caption{Plots of the normalized odd-order structure functions $-S^{||}_n(r)/(\epsilon r)^{n/3}$ vs. $r/\eta$ for a fluid simulation using Tarang.  Here $\epsilon$ is the energy flux, and $\eta$ is the Kolmogorov scale. }  
\label{fig:Str_fn}
\end{figure}

We perform code validation of the fluid solver using Kolmgorov's theory~\cite{Kolmogorov:DANS1941a} for the third-order structure function, according to which
\begin{equation}
S^{||}_3(r) = \langle \{ u_{||}({\bf x+r}) - u_{||}({\bf x}) \}^3 \rangle = -\frac{4}{5} \epsilon r
\end{equation}
where $\epsilon$ is the energy flux in the inertial range, and $\langle ... \rangle$ represents ensemble averaging (here spatial averaging).  We compute the structure function $S^{||}_3(r)$, as well as $S^{||}_5(r)$, $S^{||}_7(r)$, and $S^{||}_9(r)$ for the steady-state dataset of a fluid simulation on a $1024^3$ grid.  The computed values of $S^{||}_q(r)$ are illustrated in Fig.~\ref{fig:Str_fn} that shows a good agreement with Kolmogorov's theory.

After the discussion on fluid solver, we move on to the module for solving Rayleigh-B\'{e}nard convection.

\section{Rayleigh-B\'{e}nard convection}
Rayleigh-B\'{e}nard convection (RBC) is an idealized model of convection in which fluid is subjected between two plates that are separated by a distance $d$, and are maintained at temperatures $T_0$ and $T_0-\Delta$.  The equations for the above fluid under Boussinesq approximations are 
\begin{eqnarray}
\partial_{t}\mathbf{u} + (\mathbf{u} \cdot \nabla)\mathbf{u} & = & -\frac{\nabla \sigma}{{\rho}_0} + \alpha g \theta \hat{z} + \nu{\nabla}^2 \mathbf{u}, \label{eq:u} \\
\partial_{t}\theta + (\mathbf{u} \cdot \nabla)\theta & = & \frac{\Delta}{d} u_{z} + \kappa{\nabla}^{2}\theta, \label{eq:th} \\
\nabla \cdot {\bf u} & = & 0 \label{eq:incompressible}
\end{eqnarray}
where $\theta$ and $\sigma$ are the temperature and pressure fluctuations from the steady conduction state ($T = T_c + \theta$ with $T_c$ as the conduction temperature profile), $\hat{z}$ is the buoyancy direction, $\Delta$ is the temperature difference between the two plates,  $\nu$ is the kinematic viscosity, and $\kappa$ is the thermal diffusivity.  We solve the nondimensionalized equations, which are obtained using $d$ as the length scale, $\kappa/d$ as the velocity scale, and $\Delta$ as the temperature scale:
\begin{eqnarray}
\frac{\partial{\textbf{u}}}{\partial{t}}+ (\textbf{u}\cdot \nabla)\textbf{u} & = & -\nabla\sigma + R P \theta \hat{z} 
+  P \nabla^{2}\textbf{u},\label{eq:u_Usmall_Plarge}\\
\frac{\partial{\theta}}{\partial{t}}+(\textbf{u}\cdot\nabla)\theta  & = & u_{3} + \nabla^{2}\theta.\label{eq:T_Usmall_Plarge}
\end{eqnarray}
Here the two important nondimensional parameters are the Rayleigh number $R=\alpha g \Delta d^3/\nu \kappa$, and the Prandtl number $P=\nu/\kappa$.  In Tarang we can apply the free-slip boundary condition for the velocity fields at the horizontal plates, i.e.,
\begin{equation}
u_3= \partial_z {u_1}=\partial_z {u_2}=0 \hspace{1cm} \mathrm{for\ } z=0,1,
\end{equation}
 and  isothermal boundary condition on the horizontal plates
 \begin{equation}
\theta=0  \hspace{1cm} \mathrm{for\ } z=0,1,
\end{equation}  
Periodic boundary conditions are applied to the vertical boundaries.

The number of arrays required for a RBC simulation is 18 (15 for fluids plus three for $\theta({\bf k}), \theta({\bf r}), {\bf nlin}^\theta$).  Thus the memory requirement for RBC is (18/15) times that for the fluid simulation.  Regarding the time complexity, the number of FFT operations required per time step is $13 \times 4$ FFT operations (4 inverse + 9 forward transforms per RK4 step).  As a result, the total time requirement for a RBC simulation is (13/9) times the respective fluid simulation.

For code validation of Tarang's RBC solver, we compare the Nusselt number $Nu = 1 + \langle u_{3}\theta\rangle$ computed using Tarang with that computed by Thual~\cite{Thual:JFM1992} for two-dimensional free-slip box.  The analysis is performed for the steady-state dataset. The comparative results shown in Table~\ref{rbc_table} illustrate excellent agreement between the two runs.  We also compute the Nusselt number for a three-dimensional flow with $Pr=6.8$ and observe that $Nu = (0.27\pm 0.04)(Pr Ra)^{0.27\pm 0.01}$ ~\cite{Verma:PRE2012}, which is in good agreement with earlier experimental and numerical results.

\begin{table}[htbp]
\begin{center}
\caption{Verification of Tarang against Thual's~\cite{Thual:JFM1992} 2D
RBC simulations. We compare Nusselt numbers ($Nu$) computed in our
simulations on a $64^{2}$ grid against Thual's  simulations on $16^{2}$
(THU1), $32^{2}$ (THU2), and $64^{2}$ (THU3) grids. All $Nu$
values tabulated here are for the Prandtl number of $6.8$.}
\begin{tabular}{ c c c c c }

\\
\hline
$r$ & THU1 & THU2 & THU3 & Tarang  \\[6pt]
\hline
2 & 2.142 & -- & -- & 2.142 \\
3 & 2.678 & -- & -- & 2.678 \\
4 & 3.040 & 3.040 & -- & 3.040 \\
6 & 3.553 & 3.553 & -- & 3.553 \\
10 & 4.247 & 4.244 & -- & 4.243 \\
20 & 5.363 & 5.333 & 5.333 & 5.333 \\
30 & 6.173 & 6.105 & 6.105 & 6.105 \\
40 & 6.848 & 6.742 & 6.740 & 6.740 \\
50 & 7.441 & 7.298 & 7.295 & 7.295 \\
\hline
\end{tabular}
\label{rbc_table}
\end{center}
\end{table}

Using the RBC module of Tarang, we also studied the energy spectra and fluxes of the velocity and temperature fields~\cite{Mishra:PRE2010}, the Nusselt number scaling~\cite{Verma:PRE2012}, and chaos and bifurcations near the onset of convection~\cite{Pal:EPL2009, Paul:Chaos2011}.  

In the next section we will discuss the results of the MHD module of Tarang.

\begin{figure}[htbp]
\begin{center}
\includegraphics[scale=0.4]{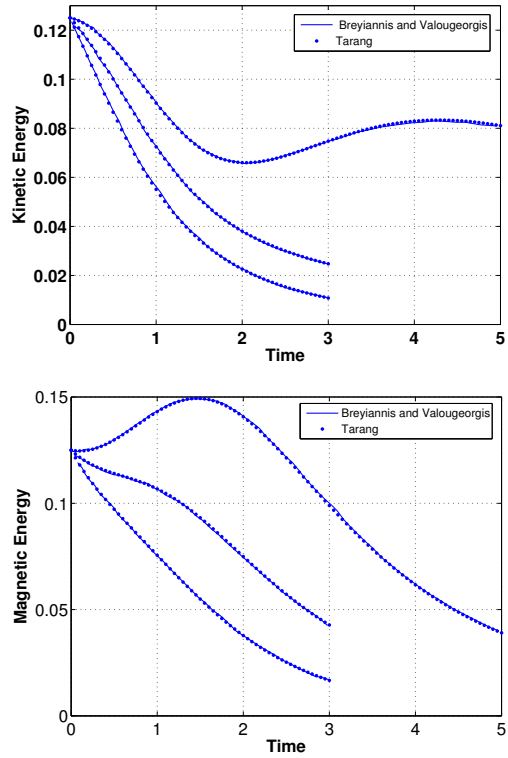}
\end{center}
\caption{Time evolution of total kinetic energy (top panel) and total magnetic energy (bottom panel) for a decaying MHD simulation with Taylor-Green vortex as an initial condition. Blue dots are Tarang's data points, while the  solid lines are the lattice simulation result of Breyiannis and Valougeorgis~\cite{Breyiannis:CF2006}. The three different curves reported here are for $\nu = \eta = 0.01$, 0.05, 0.1 from top to bottom.}  
\label{fig:KEME}
\end{figure}

\section{Magnetohydrodynamic turbulence and dynamo}
The equations for the incompressible  MHD turbulence~\cite{Verma:PR2004} are
\begin{eqnarray}
\partial_{t}{\bf u} + ({\bf u}\cdot\nabla){\bf u} &=& -\nabla{p}+  ({\bf B}\cdot\nabla){\bf B} +  \nu\nabla^2 {\bf u} + {\bf F}^u,  \quad \label{eq:MHDu}\\
\partial_{t}{\bf B} + ({\bf u}\cdot\nabla){\bf B} &=& ({\bf B}\cdot\nabla){\bf u} +  \eta \nabla^2 {\bf B}+ {\bf F}^B, \label{eq:MHDb}\\
\nabla \cdot {\bf u} = \nabla \cdot {\bf B}  & = & 0, \label{eq:MHDcontinuity} 
\end{eqnarray}
where $\bf u$, $\bf B$  and $p$ are the velocity-, magnetic-, and pressure (thermal+magnetic) fields respectively, $\nu$ is the kinematic viscosity, and $\eta$ is the magnetic diffusivity. The ${\bf F}^u$ and ${\bf F}^B$ are external forcing terms for the velocity and magnetic fields respectively. 
Typically, ${\bf F}^B = 0$, but Tarang implements ${\bf F}^B$ for generality.   The magnetic field ${\bf B}$ can be separated into its mean ${\bf B}_0$ and fluctuations ${\bf b}$: ${\bf B = B}_0+{\bf b}$.  The number of nonlinear terms in the above equations is four whose computation requires 27 FFTs.  However, the number of FFT computations in terms of the Elsasser variables ${\bf z^{\pm} = u \pm b}$ is only 15, thus saving significant computing time.  We use
\begin{eqnarray}
({\bf u}\cdot\nabla){\bf u} - ({\bf B}\cdot\nabla){\bf B} & = &
 ({\bf z^-}\cdot\nabla){\bf z^+} +  ({\bf z^+}\cdot\nabla){\bf z^-},  \\
 ({\bf u}\cdot\nabla){\bf B} - ({\bf B}\cdot\nabla){\bf u} & = &
 ({\bf z^-}\cdot\nabla){\bf z^+} -  ({\bf z^+}\cdot\nabla){\bf z^-} 
\end{eqnarray}
to compute the nonlinear terms. Thus, the time requirement for a MHD simulation would be around  15/9 times that for the fluid simulation.  In Fig.~\ref{fig:BG_Tarang_MHD_GNU} we plot the time taken per step for different set of processors  on {\em Shaheen}.  The results are consistent with the above estimates.   Regarding the space complexity, an MHD simulation requires 27 arrays for storing ${\bf u}({\bf k}), {\bf B}({\bf k}), {\bf B}({\bf r}), {\bf u}({\bf r})$, ${\bf F}^u  ({\bf k}), {\bf F}^B({\bf k}), {\bf nlin}^u({\bf k}), {\bf nlin}^B({\bf k}),$ and three temporary fields.  Hence the memory requirement for a MHD simulation is 27/15 times that of a fluid simulation.   

We perform code validation of Tarang's MHD module using the results of Breyiannis and Valougeorgis's~\cite{Breyiannis:CF2006} lattice kinetic simulations of three-dimensional decaying MHD. Following Breyiannis and Valougeorgis, we solve the MHD equations inside a cube with periodic boundary conditions on all directions, and with a  Taylor-Green vortex (given below) as an initial condition,
\begin{eqnarray}
{\bf u} & = & \left[\sin(x)\cos(y)\cos(z), -\cos(x)\sin(y)\cos(z), 0\right],\\
{\bf B} & = & \left[\sin(x)\sin(y)\cos(z), \cos(x)\cos(y)\cos(z), 0\right].
\end{eqnarray}
This Taylor-Green vortex is then allowed to evolve freely. The simulation box is discretized using $32^3$ grid points.

The results of this test case for different parameter values ($\nu = \eta = 0.01$, 0.05, 0.1) are presented in Fig.~\ref{fig:KEME}. The top and bottom panels exhibit the time evolution of the total kinetic- and magnetic energies respectively.  Tarang's data points, illustrated using blue dots, are in excellent agreement with Breyiannis and Valougeorgis' results~\cite{Breyiannis:CF2006}, which is represented using solid lines. We thus verify the MHD module of Tarang.

We have used Tarang to perform extensive simulations of dynamo transition under the Taylor-Green forcing~\cite{Yadav:EPL2010, Yadav:PRE2012}.   Using Tarang, we have also computed the magnetic and kinetic energy spectra, various energy fluxes~\cite{Verma:PR2004}, and shell-to-shell energy transfers for MHD turbulence; these results would be presented in a subsequent paper.

In addition to the fluid, MHD, and Rayleigh-B\'{e}nard convection solvers, Tarang has modules for simulating rotating turbulence, passive and active scalars, liquid metal flows, rotating convection~\cite{Pharsi:PRE2011}, and Kolmogorov flow. 

\section{Conclusions}
In this paper we describe salient features and code validation of Tarang.  Tarang passes several validation tests performed for fluid, Rayleigh-B\'{e}nard convection, and  magnetohydrodynamic solvers.  We also report scaling analysis of Tarang and show that it exhibits excellent strong- and weak  scaling up to several thousand processors.     Tarang has been used for studying Rayleigh-B\'{e}nard convection, dynamo, and magnetohydrodynamic turbulence.  It has been ported to various computing platforms including the {\em HPC system} of IIT Kanpur, {\em Shaheen} of KAUST, {\em Param Yuva} of the Centre for Advanced Computing (Pune), and {\em EKA} of the Computational Research Laboratory (Pune).

\section*{Acknowledgement}
Tarang simulations were performed on {\em Shaheen} supercomputer of KAUST (through the project k97) and on the {\em HPC system} of IIT Kanpur, for which we thank the personnels of respective Supercomputing Centers, especially Abhishek and Brajesh Pande of IIT Kanpur.  We are grateful to Sandeep Joshi and Late Dr. V. Sunderarajan (CDAC) who encouraged us to run Tarang on very large grids.  We also thank Daniele Carati and his group at ULB Brussels for sharing with us the details of a pseudospectral code, and CDAC engineers for help at various stages.  MKV acknowledges the support of Swaranajayanti fellowship and a research grant 2009/36/81-BRNS from Bhabha Atomic Research Center.
\bibliographystyle{model3-num-names}

\end{document}